\documentclass[aps,showpacs,prl,twocolumn,floatfix]{revtex4}

\usepackage[english]{babel}
\usepackage{subfigure}
\usepackage{latexsym}
\usepackage{graphicx}
\usepackage{epsfig}

\begin{document}

\title {Exact Stochastic Reformulation of Non-Relativistic Quantum Dynamics}
\author{Murat \c{C}etinba\c{s} and Joshua Wilkie}
\affiliation{Department of Chemistry, Simon Fraser University, Burnaby,
             British Columbia V5A 1S6, Canada}

\begin{abstract}
We show that any non-relativistic quantum $N$-body dynamics problem with pairwise interactions can be exactly 
reformulated in terms of $N$ well behaved 1-body stochastic density equations. Specifically, the time evolving
$N$-body density matrix is written as an average of tensor products of 1-body densities each of which
obeys a stochastic evolution equation. Such decompositions can be constructed for any mixture of fermions and bosons. The evolution equations for the 1-body densities preserve norm, Hermiticity and positivity.
\end{abstract}


\maketitle


While the dynamical laws underlying much of chemistry and physics are known, these
equations can only rarely be solved analytically, and in fact cannot even be solved numerically except when the number of
participating particles is very small. This is a major difficulty for theoretical chemistry and solid state physics where
systems of many electrons must be solved in order to determine electronic structure. Approximation methods such as density 
functional theory and Hartree-Fock and its generalizations allow estimates of ground state energies and other properties, but limited accuracy and generality is inherent in such approaches. Other more formal computational approaches exist, which are based on Lanczos type algorithms, configuration interaction\cite{Sch} and coupled-cluster\cite{Bart} schemes, which converge {\em in principle} to exact solutions in appropriate limits. The difficulty of such methods stems from the 
practical necessity of representing the wavefunction in a basis of 1-body states\cite{GG}. That is, if $m$ basis functions are required for each body then $m^N$ complex numbers will be needed to represent the wavefunction. This exponential growth of computational complexity with $N$ is characteristic of almost all numerical strategies which converge to exact solutions as $m\rightarrow\infty$. Very recently several classes of computational methods have been discovered which appear to circumvent this prohibitive scaling\cite{GG,Koonin,CCD,JC,TWC}.

The methods of interest here are those based on wave equations and stochastic representation of the pairwise interaction\cite{CCD,JC,TWC}. Polynomially scaling schemes for evolution of systems of identical bosons\cite{CCD} and fermions\cite{JC} have been 
developed which decompose the state vector as an average of solutions of 1-body stochastic wave equations. These methods
avoid explicit construction or storage of the state vector and require storage which scales only linearly with $N$. Actual
computational costs scale as some small power of $N$. Unphysical properties of such decompositions, such as exponential growth of 1-body norm\cite{CCD,JC}, have been eliminated by rederivation of the evolution equations\cite{TWC} from a variational principle\cite{W1}. In addition, considerable progress has been made in developing numerical strategies for solving the 1-body
stochastic evolution equations\cite{W3}. These developments suggest that stochastic reformulations of the $N$-body problem 
may ultimately lead to more effective computational methods for quantum chemistry and physics.

An interesting open question is whether it is possible to reformulate {\em all} non-relativistic many body problems using such stochastic decompositions? Existing decompositions work only for systems of identical fermions {\em or} systems of identical bosons. Mixtures are not allowed. These restrictions on composition seem artificial and limit application. Fundamental physical systems as simple as heteronuclear molecules cannot be evolved with these methods. In this manuscript we develop a general unified theory applicable to any non-relativistic $N$-body system of pairwise interacting distinguishable and indistinguishable quantum particles. This is achieved by expressing the $N$-body density matrix as an average of tensor products of $N$ 1-body densities each of which obeys a stochastic evolution equation. We also show that the stochastic evolution equations are surprisingly well behaved with preservation of 1-body norm, Hermiticity and positivity. Our variational derivation also shows that the decomposition with these favorable properties is unique. Finally, we provide a simple recipe for recovery of the state vector from the $N$-body density matrix.

Consider then the general Hamiltonian for a system of particles with pairwise interactions
\begin{eqnarray}
{\cal H}_{N} = \sum_{k=1}^{N} H_k + \sum_{k=1}^{N-1} \sum_{l=k+1}^{N} V_{k,l}
\label{H} 
\end{eqnarray}
where $H_k$ denotes the single body Hamiltonian for particle $k$ and $V_{k,l}$ is a general interaction potential for particles $k$ and $l$. Without loss of generality we write the general 2-body interaction $V_{k,l}$ as a sum of products of 1-body operators
\begin{eqnarray}
V_{k,l} = \sum_{s=1}^{p} \omega_{s} O_k^s O_l^s
\label{V} 
\end{eqnarray}
where $p$ may be arbitrarily large. [Note that for a Coulomb interaction $p$ is in principle infinite, but in practice restriction to a finite basis of 1-body states leads to finite $p$.] This decomposition can {\em always} be chosen such that the dimensionless 1-body operators $O_k^s$ are Hermitian. The real coefficients $\omega_{s}$ have units of energy and may be positive or negative. 

Leaving issues of symmetry due to indistinguishability for later, consider an initial $N$-body density matrix which is a product of 1-body densities. Exact dynamics of the N-body density $\rho^{N} (t)$ will be governed by the Liouville-von Neumann equation. We will look for a representation of the time evolved density as an average tensor product of 1-body densities. Specifically, we will derive a decomposition of the form
\begin{equation}
\rho^{N} (t)={\rm M}[\rho_1 (t)\dots \rho_N(t)]
\end{equation}
where $\rho_k (t)$ for $k=1, \ldots , N$ are intended to be one-body density operators which obey stochastic evolution equations yet to be determined, and ${\rm M}[\dots ]$ denotes an average over different realizations of the stochastic processes. Our
decision to decompose the density matrix rather than the state vector is motivated by a proof\cite{W4} that no such {\em general} decomposition conserves 1-body norm.

Introducing a set of complex Wiener processes $\alpha_{k,l}^s(t)$ the most general form of the stochastic evolution 
equations is
\begin{eqnarray}
d\rho_k(t)&=& v_k(t)  dt + \sum_{s=1}^{p} u_k^s (t)\sum_{l \neq k}^{N} d\alpha_{k,l}^{s}(t) \nonumber \\
&+& \sum_{s=1}^{p} u_k^{s\: \dag}(t) \sum_{l \neq k}^{N} d\alpha_{k,l}^{s\: \ast}(t),
\label{decomp} 
\end{eqnarray}
where $v_k(t)$ and $u_k^s (t)$ are unknown operators to be determined, and
where $v_k(t)$ must be Hermitian in order that Hermiticity be preserved. The complex stochastic differentials $d\alpha_{k,l}^{s}(t)$ are of the form $(\mu+i\nu)\sqrt{dt}$ where $\mu,\nu$ are normally distributed with zero mean and unit variance. Additionally, we require that ${\rm Tr}_{k}\{v_k\}=0$ and ${\rm Tr}_{k} \{ u_k^s \}=0$ so that trace norm will be conserved. If we employ It\^{o} calculus\cite{Gard} then
\begin{eqnarray}
d\rho^{N}={\rm M}[\sum_{k=1}^{N} d\rho_k \prod_{l \neq k}^{N} \rho_l + \! \sum_{k=1}^{N-1} \sum_{l=k+1}^{N}\!\! d\rho_k d\rho_l \! \! \prod_{m \neq k,l}^{N}\!\! \rho_m ]
\label{drho}
\end{eqnarray}
and we will require that this equal $-i[{\cal H}_{N},\rho^{N}]dt$ so that our decomposition is exact (note that in our units $\hbar=1$). That is, we require that the average tensor product equal the exact $N$-body density matrix $\rho^N$ which obeys the Liouville-von Neumann equation $d\rho^{N}=-i{\cal L} \rho^{N}dt$, ${\cal L}$ being the Liouville super-operator ${\cal L} = [{\cal H}_{N},\, .\;]$. Substitution of (\ref{decomp}) into (\ref{drho}) reveals that we must have terms in $u_k^s$ which are proportional to $O_k^s$
in order that the pairwise coupling terms in ${\cal H}_{N}$ be reproduced. In addition to avoid unwanted terms it will be necessary to require that
$d\alpha_{l,k}^{s}=d\alpha_{k,l}^{s\: \ast}$ and that the Wiener processes be otherwise independent. Standard rules of It\^{o} calculus\cite{Gard} then determine the following further relations
\begin{equation}
d\alpha_{k,l}^{s\: \ast}d\alpha_{k',l'}^{s'}=\delta_{k,k'}\delta_{l,l'}\delta_{s,s'}dt\label{cons}
\end{equation}
among stochastic differentials.

To find explicit forms for $v_k$ and $u_k^s$ by trial and error would be difficult. The task is greatly simplified by use of
a variational principle\cite{W1}. Defining a Hilbert-Schmidt norm $\| x \|^2=(x|x)$ and inner product $(x|y) = {\rm Tr}\{x^{\dag} y\}$ (see \cite{W2}) we may express the mean square error in a decomposition as a functional
\begin{equation}
{\cal F} = \parallel d\rho^{N} + i{\cal L} \rho^{N}dt\parallel^{2}
\label{func}
\end{equation}
which can be minimized by variation of $v_k$ and $u_k^s$. Allowing all independent variations $\delta v_k,\; \delta u_k^s$, and $\delta u_k^{s\: \dag}$, substituting Eqs. (\ref{drho}) and (\ref{decomp}) into Eq. (\ref{func}), keeping only terms of order $dt$ and dropping $dt$ as a factor, then gives
\begin{eqnarray}
0&=&\delta {\cal F}={\rm Tr} \{ ( \sum_{k=1}^{N} \delta v_k \prod_{l \neq k}^{N} \rho_l  \nonumber \\
&+& \sum_{k=1}^{N-1} \sum_{l=k+1}^{N} \sum_{s=1}^{p} 
\delta [ u_k^s u_l^s + u_k^{s\: \dag} u_l^{s\: \dag} ] \prod_{m \neq k,l}^{N} \rho_m  |  \nonumber \\
& &  \sum_{k=1}^{N} v_k \prod_{l \neq k}^{N}\rho_l + \sum_{k=1}^{N-1} \sum_{l=k+1}^{N} \sum_{s=1}^{p} [u_k^s u_l^s + u_k^{s\: \dag}u_l^{s\: \dag}]
\prod_{m \neq k,l}^{N} \rho_m \nonumber \\
&+&i\ {\cal H}_{N}\rho^{N}-i\ \rho^{N}{\cal H}_{N} ) \} 
\label{dfunc} 
\end{eqnarray} 
where $\delta [u_k^s u_l^s + u_k^{s\: \dag} u_l^{s\: \dag}]=\delta u_k^s u_l^s +u_k^s \delta u_l^s +\delta u_k^{s\: \dag}u_l^{s\: \dag}+u_k^{s\: \dag}\delta u_l^{s\: \dag}$. 
The solution of Eq. (\ref{dfunc}) still involves tedious algebra and hence the complete derivation will be given elsewhere\cite{W4}. One finds that
\begin{eqnarray}
d\rho_k(t)&=&-i\left[ H_k,\rho_k(t) \right]\!~dt -i  \sum_{s=1}^{p} \sum_{l \neq k}^{N}\omega_s \overline{O}_l^s(t) \left[ O_k^s,\rho_k (t)\right]\!~dt \nonumber \\ 
&+&  \sum_{s=1}^{p}\sqrt{-i\omega_s} ( O_k^s - \overline{O}_k^s(t) ) \rho_k(t) \; \sum_{l \neq k}^{N}  d\alpha_{k,l}^{s} (t)
\; \nonumber \\
&+& \;  \sum_{s=1}^{p}(\sqrt{-i\omega_s})^* \rho_k(t) ( O_{k}^{s} - \overline{O}_{k}^{s}(t) ) \; \sum_{l \neq k}^{N} d\alpha_{k,l}^{s\: \ast}(t)
\label{decomp2} 
\end{eqnarray}
where $\overline{O}_k^s(t) = {\rm Tr}_{k} \{ O_k^s \rho_k (t)\}$. A detailed examination of this derivation indicates that these equations are unique\cite{W4}. We now focus on the properties of these evolution equations for the stochastic one-body density operators.

{\em Proof of exactness}: Substituting Eq. (\ref{decomp2}) into (\ref{drho}), using Eqs. (\ref{cons}), recognising that the differentials are statistically independent of the densities at time $t$, and using the fact that $M[d\alpha_{k,l}^{s}]=0$ we have
\begin{eqnarray}
& &d\rho^{N}={\rm M} [-i\sum_{k=1}^{N}\;[\: H_k,\rho_k \:]\prod_{l\neq k}^{N}\rho_l \nonumber \\
& &-i \sum_{k=1}^{N}\sum_{l\neq k}^{N}\sum_{s=1}^{p}\omega_s \overline{O}_l^s\; 
[\: O_k^s, \rho_k \:] \prod_{l \neq k}^{N} \rho_l  \\ 
& & -i \sum_{k=1}^{N-1}\sum_{l=k+1}^{N}\sum_{s=1}^{p}\omega_s ( O_k^s -\overline{O}_k^s )( O_l^s -\overline{O}_l^s ) \:\rho_k \rho_l \!\! \prod_{m \neq k,l}^{N}\!\! \rho_m \nonumber \\ 
& &+i \sum_{k=1}^{N-1} \sum_{l=k+1}^{N} \sum_{s=1}^{p}\omega_s \rho_k \rho_l \: ( O_k^s -\overline{O}_k^s )(O_l^s -\overline{O}_l^s )\!\! \prod_{m \neq k,l}^{N} \!\!\rho_m ]\!\!~dt\nonumber
\end{eqnarray}
By definition $\rho^N={\rm M}[\rho_k \prod_{l \neq k}^{N}\rho_l]={\rm M} [\rho_k \rho_l \prod_{m \neq k,l}^{N}\rho_m ]$, and so after some algebra it follows that
\begin{eqnarray}
d\rho^{N}&=&-i\sum_{k=1}^{N} [\: H_k,\rho^{N}\:]~dt \nonumber \\
&-&i \sum_{k=1}^{N-1}\sum_{l=k+1}^{N}\sum_{s=1}^{p}\omega_s [\:O_k^s O_l^s, \rho^{N}\:]~dt  \nonumber \\
&-&i \sum_{k=1}^{N}\sum_{l \neq k}^{N} \sum_{s=1}^{p} \omega_s \overline{O}_l^s \;[\:O_k^s , \rho^{N}\:]~dt \label{ext2}\\
&+& i  \sum_{k=1}^{N-1}\sum_{l=k+1}^{N}\sum_{s=1}^{p}\omega_s \left\{ \overline{O}_l^s [ \:O_k^s ,\rho^{N}\:]+\overline{O}_k^s [\:O_l^s ,\rho^{N}\:] \right\}~dt.\nonumber 
\end{eqnarray}
Finally, using the following relation of the summation operators in Eq. (\ref{ext2})
\begin{eqnarray}
\sum_{k=1}^{N}\sum_{l \neq k}^{N}[\:.\:]_k [\:.\:]_l = \sum_{k=1}^{N-1}\sum_{l=k+1}^{N} [\:.\:]_k[\:.\:]_l+[\:.\:]_l[\:.\:]_k \nonumber
\end{eqnarray}
the last two terms in Eq. (\ref{ext2}) cancel out and we obtain the Liouville-von Neumann equation
\begin{eqnarray}
d\rho^{N}&=&-i \left[\: \sum_{k=1}^{N} H_k + \sum_{k=1}^{N-1}\sum_{l=k+1}^{N} \sum_{s=1}^{p} \omega_s O_k^s O_l^s ,\;\rho^{N} \: \right]~dt \nonumber \\
&=&-i\left[\: {\cal H}_{N},\rho^{N}\:\right]~dt
\end{eqnarray}
\noindent
which proves that the decomposition is exact.
 
{\em Proof of Hermiticity}: To show $\rho_k (t)= \rho_k^{\dag}(t)$ take the adjoint of Eq. (\ref{decomp2}). Since this results in the same equation with the same initial condition it follows that $\rho_k (t)$ is Hermitian.

{\em Proof of norm conservation}: In order to prove that trace-norm is preserved we trace over Eq. (\ref{decomp2}) which yields
\begin{eqnarray*}
{\rm Tr}_{k} \{ d\rho_k \} &=& -i {\rm Tr}_{k} \{ [ H_k ,\rho_k ] \} \!~dt \nonumber \\
&-&i  \sum_{s=1}^{p} \sum_{l \neq k}^{N}\omega_s \overline{O}_l^s {\rm Tr}_{k} \{ [ O_k^s ,\rho_k ] \} \!~dt \nonumber \\ 
&+& \sum_{s=1}^{p}\sqrt{-i\omega_s} {\rm Tr}_{k} \{ ( O_k^s -\overline{O}_k^s ) \rho_k \} \sum_{l \neq k}^{N} d\alpha_{k,l}^{s} \nonumber \\
&+& \sum_{s=1}^{p} (\sqrt{-i\omega_s})^* {\rm Tr}_{k} \{ \rho_k ( O_k^s -\overline{O}_k^s ) \} \sum_{l \neq k}^{N} d\alpha_{k,l}^{s\: \ast} \nonumber \\
& =&0
\end{eqnarray*}
where we have used the fact that ${\rm Tr}\{AB\}={\rm Tr}\{BA\}$ and the definition of $\overline{O}_k^s$.
 
{\em Proof of positivity}: Surprisingly, the one-body density operators remain positive semi-definite. To see this
recall that the Hermitian property of the stochastic one-body density operators permits a spectral decomposition of the form
\begin{equation}
\rho_k (t)=\sum_{j=1}^{\infty} w_{k}^{j}(t){\cal E}_{k}^{j}(t)\label{E1}
\end{equation}
where ${\cal E}_{k}^{j}(t)=|k,j;(t)\rangle \langle k,j;(t)|$ are the spectral projection operators for the natural orbitals and $w_{k}^{j}(t)$ are their occupation 
probabilities (i.e. $\rho_k (t) |k,j;(t)\rangle =w_{k}^{j}(t) |k,j;(t)\rangle$). The question is whether the $w_{k}^{j}(t)$ are positive for all time.
Hence, consider the following lemma:

{\em Lemma}: The decomposition (\ref{E1}) satisfying Eqs. (\ref{decomp2}) implies that
\begin{eqnarray}
dw_{k}^{j}(t)&=& \sum_{s=1}^{p}\sqrt{-i\omega_s } ( \langle O_{k}^{s} \rangle_j - \overline{O}_{k}^{s} ) w_{k}^{j}(t) \sum_{l \neq k}^{N} d\alpha_{k,l}^{s}\nonumber \\
&+& \sum_{s=1}^{p}(\sqrt{-i\omega_s })^* ( \langle O_{k}^{s} \rangle_j - \overline{O}_{k}^{s} ) w_{k}^{j}(t) \sum_{l \neq k}^{N} d\alpha_{k,l}^{s\: \ast}
 \label{DYN1} \\ 
d{\cal E}_{k}^{j}(t)&=&-i \left[ H_k, {\cal E}_{k}^{j}(t) \right]dt -i \sum_{s=1}^{p}\sum_{l \neq k}^{N} \omega_s \overline{O}_l^s \left[O_k^s,{\cal E}_{k}^{j}(t)\right]dt
\nonumber \\
&-& (N-1)\sum_{s=1}^{p}\omega_s  ( \langle O_{k}^{s} \rangle_j -\overline{O}_{k}^{s})(O_{k}^{s}-\langle O_{k}^{s} \rangle_j ){\cal E}_{k}^{j}(t)~dt\nonumber \\
&-& (N-1)\sum_{s=1}^{p}\omega_s  {\cal E}_{k}^{j}(t)( \langle O_{k}^{s} \rangle_j - \overline{O}_{k}^{s} )( O_{k}^{s} - \langle {O}_{k}^{s} \rangle_j )~dt \nonumber \\
&+& \sum_{s=1}^{p}\sqrt{-i\omega_s } ( O_{k}^{s} - \langle {O}_{k}^{s} \rangle_j ) {\cal E}_{k}^{j}(t) \sum_{l \neq k}^{N}d\alpha_{k,l}^{s}\nonumber \\
&+& \sum_{s=1}^{p}(\sqrt{-i\omega_s})^* {\cal E}_{k}^{j}(t) ( O_{k}^{s}- \langle {O}_{k}^{s}\rangle_j ) \sum_{l \neq k}^{N} d\alpha_{k,l}^{s\: \ast}.
\label{DYN2}
\end{eqnarray}
\noindent
Here $ \langle O_{k}^{s}\rangle_j ={\rm Tr}_{k}\{O_k^s {\cal E}_{k}^{j}(t)\}$.

{\em Proof}: From (\ref{E1}) it follows that
\begin{eqnarray}
d\rho_k&=&\sum_{j=1}^{\infty} dw_k^j(t){\cal E}_{k}^{j}(t)+\sum_{j=1}^{\infty} w_k^j(t)d{\cal E}_{k}^{j}(t)\nonumber \\
&+&\sum_{j=1}^{\infty} dw_k^j(t) d{\cal E}_{k}^{j}(t)
\end{eqnarray}
and substituting Eqs. (\ref{DYN1}) and (\ref{DYN2}) and using the relations (\ref{cons}) then gives (\ref{decomp2}). 

Finally, from Eq. (\ref{DYN1}) it readily follows that the occupation probabilities must satisfy the relation
\begin{eqnarray}
& &w_k^j(t)=\prod_{s=1}^{p} \prod_{l \neq k}^{N} \exp \{ \nonumber \\
& &2\int_0^t ( \langle O_{k}^{s} \rangle_j(t') - \overline{O}_{k}^{s}(t') ) {\rm Re}[ \sqrt{-i\omega_s} d\alpha_{k,l}^{s}(t')]\nonumber \\
& & - \omega_s \int_0^t ( \langle O_{k}^{s} \rangle_j(t') - \overline{O}_{k}^{s}(t') )^2 dt' \}\nonumber \\
& &\geq  0
\end{eqnarray}
which shows that the occupation probabilities are indeed positive. It follows that the 1-body densities are positive semi-definite.

Finally, we outline a simple method to extract the wave vector from the density operator. Suppose the initial state for the $N$-body system is a pure state $\Psi^{N}(0)$. Now choose a set of arbitrary states $|i_k\rangle$, one for each particle. Then construct
\begin{eqnarray}
|\tilde{\Phi}^{N} (t) \rangle = {\rm M} \left[ \rho_1 (t)|i_1 \rangle \otimes \rho_2 (t)|i_2 \rangle \otimes\ldots \otimes \rho_N(t)|i_N\rangle \right]
\label{wave}
\end{eqnarray}
and normalize to obtain $|\Phi^{N}(t)\rangle=|\tilde{\Phi}^{N}(t)\rangle / \| \tilde{\Phi}^{N}(t)\|$. This vector now differs
from the exact wave function by a possible phase factor: that is
\begin{eqnarray}
|\Phi^{N} (t)\rangle = |\Psi^{N}(t)\rangle \exp [i\Theta (t)]
\label{ph}  
\end{eqnarray}
with $d|\Psi^{N}(t)\rangle=-i{\cal H}_N |\Psi^{N}(t)\rangle dt$. It can be readily shown that
\begin{eqnarray}
\Theta (t)= \int_{0}^{t} \frac{ \langle \Psi^{N}(0)|{\cal H}_{N} |\Phi^{N}(t')\rangle - i \frac{ \partial}{\partial t'} \langle \Psi^{N}(0)|\Phi^{N}(t') \rangle }
{ \langle \Psi^{N}(0)|\Phi^{N}(t')\rangle }~dt'
\label{ph3}
\end{eqnarray}
where the derivative with respect to $t'$ is best computed via fast Fourier transform in practice. Since we have a closed formula for the phase the exact wave function for distinguishable particles reads
\begin{eqnarray}
|\Psi^{N}(t)\rangle = |\Phi^{N}(t)\rangle \exp [-i\Theta (t)]
\label{ph4}  
\end{eqnarray}
Note that the result (\ref{ph4}) is itself a sum of tensor products.
From this expression one may then impose appropriate symmetries for indistinguishable particles by operating with symmetrization and anti-symmetrization operators. Thus the wave vector can be recovered from the density operator and hence it is possible to compute all quantum mechanical quantities using our approach. It is worth noting that average quantities such as $\bar{A}={\rm Tr}\{A\rho^N\}$ are easily computed with our equations but cannot be easily obtained with other approaches\cite{CCD,JC,TWC}.
Eigenspectra are readily calculated by the Fourier transform of the autocorrelation function $\langle \Psi^{N}(0)| \Psi^{N}(t)\rangle$ using the formula
\begin{eqnarray}
I(E)&=&\langle \Psi^{N}(0)|\delta (E-{\cal H}_{N})|\Psi^{N}(0)\rangle \nonumber \\
&\simeq &\frac{1}{\pi\hbar} {\rm Re} \int_{0}^{T} 
\langle \Psi^{N}(0)|\Psi^{N}(t)\rangle
\exp \left( iEt \right) \; dt
\label{spect}
\end{eqnarray}
for $T$ sufficiently large.

In conclusion, we have developed an exact and completely general reformulation of non-relativistic quantum dynamics 
for pairwise interacting particles by decomposing the density operator of the $N$-body problem as an average of tensor products of stochastic 1-body density operators. We showed that during the stochastic evolution the norm, Hermiticity, and the positivity of the stochastic 1-body density operators are exactly preserved. Finally we explained how the wave vector and hence all
quantum mechanical quantities of interest can be recovered. A longer manuscript which presents details of the derivation
and proves uniqueness of the decomposition is in preparation\cite{W4}. Our equations may have applications as a computational 
technique for chemistry and solid state problems, and might also prove useful in themselves as a theoretical tool for predicting quantities
such as fluctuations in single molecule fluorescence\cite{BJS}. 

The authors acknowledge the financial support of the Natural
Sciences and Engineering Research Council of Canada.

\end{document}